\RequirePackage{fix-cm}
\documentclass[smallextended]{svjour3}       % onecolumn (second format)
\smartqed  % flush right qed marks, e.g. at end of proof
\usepackage{graphicx}
\usepackage{natbib}
\usepackage{xcolor}
\usepackage[utf8]{inputenc}
\usepackage{amssymb}
\usepackage{rotating}

\begin{document}

\title{Is there a universal parametric city size distribution? Empirical evidence for 70 countries
\thanks{Financial support from Ministerio de Economía y Competitividad (ECO2017-82246-P) and support by Aragon Government (ADETRE Consolidated Group) is acknowledged.}
}

\titlerunning{Is there a universal parametric city size distribution?}        % if too long for running head

\author{Miguel Puente-Ajov\'\i{}n    \and
        Arturo Ramos            \and
        Fernando Sanz-Gracia
}

\authorrunning{M. Puente-Ajov\'\i{}n et al.} % if too long for running head

\institute{M. Puente-Ajovín (corresponding author) \at
              Department of Economic Analysis \\
              University of Zaragoza (Spain)\\
              \email{mpajovin@unizar.es}           %  \\
           \and
           A. Ramos \at
           Department of Economic Analysis \\
              University of Zaragoza (Spain)\\
              \email{aramos@unizar.es}
           \and
           F. Sanz-Gracia \at
           Department of Economic Analysis \\
              University of Zaragoza (Spain)\\
              \email{fsanz@unizar.es}
}

\date{Received: date / Accepted: date}

\maketitle
\begin{abstract}
We study the parametric description of the city size distribution (CSD) of 70 different countries (developed and developing) using seven models, as follows: the lognormal (LN), the loglogistic (LL), the double Pareto lognormal (dPLN), the two-lognormal (2LN), the two-loglogistic (2LL), the three-lognormal (3LN) and the three-loglogistic (3LL). Our results show that  3LN and 3LL are the best densities in terms of non-rejections out of standard statistical tests. Meanwhile, according to the information criteria AIC and BIC, there is no systematically dominant distribution.
\end{abstract}

\section{Introduction}
\label{intro}
The study of city size distributions has a long tradition in urban economics. Following \cite{MalPisSor11}, this is an interesting research field for the following reasons: first, the study of the shape of the distribution can give information about its underlying growth mechanisms and processes; second, it yields important socioeconomic consequences; third, the upper tail is, by definition, very relevant in terms of population; and fourth, the better we can describe a distribution, the better we can predict its future shape and anticipate behaviors to come. This is especially important for city size distribution given that the world is becoming increasingly urban and ``the future of humankind will be an urban one'' \citep{mcgranahan2014urban}. In this context, according to the \cite{northam1979urban} curve, once a country reaches an urbanization rate of 30\%, a fast and accelerated urbanization stage follows.
\newline
\newline
Chronologically, the first density function to be considered, especially for the upper tail, was the Pareto distribution and a particular case of it known as Zipf’s distribution \citep{rosenResnick80}. \cite{parr1973spa} proposed the lognormal density and \cite{eeckhout2004gsl} developed a theoretical model that was based on it. Trying to conciliate these two approaches, \cite{reed2004double}, theoretically, and \cite{giesen2010size} and \cite{val15}, empirically, defend the so-called double Pareto lognormal (dPLN), featuring Pareto tails convolved with a lognormal body. In this context, \cite{PueRam15} consider the best model for some European countries to have Pareto tails and a Singh--Maddala body (tdPSM). Subsequently, \cite{lude17} propose a variant of the dPLN (Pareto tails and lognormal body, PTLN) which performs better than the dPLN. Later on, \cite{kwna19} introduced finite mixtures of lognormal distributions (or, more generally, log-exponential power distributions, LEPs) in the study of city size distributions, which outperform the  PTLN for India and the United States. Afterwards, finite mixtures of lognormal distributions have also been used by
\cite{BanChiPrePueRam19} for Romania and \cite{su19} for the United States.
\newline
\newline
Against this background, in this study we try to test whether the result of \cite{kwna19} is applicable to other countries. To that end and to give robustness to the conclusions, we have considered a wide sample of 70 countries  (see Section~\ref{Data}). We have estimated seven distributions, as follows: the lognormal (LN), the loglogistic (LL), the dPLN, mixtures of two and three lognormals (2LN and 3LN) and mixtures of two and three loglogistics (2LL and 3LL).
The 3LN is the most representative distribution proposed by \cite{kwna19} and the LL was first proposed in \cite{Hsu12} (see also \cite{val15}) to study city size distributions, although it is widely used in other branches of economics \citep{Fis61}. As far as we know, 2LL and the 3LL are
used here for the first time to describe city size distributions.
\newline
\newline
The rest of the paper is organized as follows. Section~\ref{functions} describe the density functions that are considered in this study. Section~\ref{Data} describes the choice and type of dataset. In Section~\ref{Results}, we present the empirical results. Finally, we draw our conclusions in Section~\ref{conclusions}.

\section{The density functions}
\label{functions}

The well-known lognormal (LN) density function depends on two parameters $\mu,\sigma>0$. It is defined for the support $x>0$ and its expression is given by
$$
f_{{\rm LN}}(x,\mu,\sigma)=\frac{1}{x\sigma\sqrt{2\pi}}
\exp{\left(-\frac{(\ln(x)-\mu)^2}{2\sigma^2}\right)}
$$

Similarly, the loglogistic (LL) also depends on two parameters, also denoted by $\mu,\sigma>0$ with the same support $x>0$ and can be expressed as
$$
f_{{\rm LL}}(x,\mu,\sigma)=\frac{\exp{\left(-\frac{\ln(x)-\mu}{\sigma}\right)}}
{x\sigma\left(1+\exp{\left(-\frac{\ln(x)-\mu}{\sigma}\right)}\right)^2}
$$

The third distribution in our study is
the double Pareto lognormal distribution (dPLN), which was
introduced by \cite{Ree02,Ree03}, where the density function reads:
\begin{eqnarray}
f_{\rm dPLN}(x,\alpha,\beta,\mu,\sigma)
&=&\frac{\alpha\beta}{2x(\alpha+\beta)}
\exp\left(\alpha\mu+\frac{\alpha^{2}\sigma^{2}}{2}\right)x^{-\alpha}
\left(1+{\rm erf}\left(\frac{\ln(x)-\mu-\alpha\sigma^{2}}
{\sqrt{2}\sigma}\right)\right) \nonumber\\
& &-\frac{\alpha\beta}{2x(\alpha+\beta)}
\exp\left(-\beta\mu+\frac{\beta^{2}\sigma^{2}}{2}\right)x^{\beta}
\left({\rm erf}\left(\frac{\ln(x)-
\mu+\beta\sigma^{2}}{\sqrt{2}\sigma}\right)-1\right)\nonumber
\end{eqnarray}
where ${\rm erf}$ is the error function associated to the normal
distribution and the four
parameters of the
distribution are $\alpha,\beta,\mu,\sigma>0$. This has the property that it approximates different power laws in each of its two tails:
$f_{\rm dPLN}(x)\approx x^{-\alpha-1} $ when $x\to\infty$
and
$f_{\rm dPLN}(x)\approx x^{\beta-1}$ when $x\to 0$,
hence the name double Pareto. The body is approximately
lognormal, although it is not possible to exactly delineate
the switch between the lognormal and the Pareto behaviors.

The finite mixtures 2LN and 3LN \citep{kwna19} are a convex linear combination of $2$ or $3$ lognormals whose densities can be written as
$$
f_{{\rm 2LN}}
(x,\mu_1,\sigma_1,\mu_2,\sigma_2,p_1)=p_1 f_{{\rm LN}}(x,\mu_1,\sigma_1)+(1-p_1)f_{{\rm LN}}(x,\mu_2,\sigma_2)
$$
where $0\leq p_1\leq 1$, $0\leq 1-p_1\leq 1$, and $\mu_i,\sigma_i>0$, $i=1,2$, and
$$
f_{{\rm 3LN}}
(x,\mu_1,\sigma_1,\mu_2,\sigma_2,\mu_3,\sigma_3,p_1,p_2)=p_1 f_{{\rm LN}}(x,\mu_1,\sigma_1)+p_2 f_{{\rm LN}}(x,\mu_2,\sigma_2)+(1-p_1-p_2)f_{{\rm LN}}(x,\mu_3,\sigma_3)
$$
where $0\leq p_1,p_2\leq 1$, $0\leq 1-p_1-p_2\leq 1$, and $\mu_i,\sigma_i>0$, $i=1,2,3$.
The 2LL and 3LL are defined in a similar way by everywhere replacing $f_{{\rm LN}}$ by $f_{{\rm LL}}$.

In this article, for the sake of simplicity and convergence of the estimations, we will consider only  2LN, 3LN, 2LL and 3LL  because, as a general rule, we hold that when the number of components in the mixture is higher, the convergence of Maximum Likelihood Estimation (MLE) of the parameters of the distribution will be  slower.

\section{Data}
\label{Data}
The aim of this paper is to represent the majority of the city size distributions around the globe. Unfortunately, not all countries make their population census available. Consequently, we decided to select 70 countries that represent, as closely as possible, the diversity of countries in the world. To do so, we use the following five variables to account for this heterogeneity: Human Development Index, GDP per capita, international trade (the sum of imports and exports over GDP), geographical area, and total population. This allows us to take the level of development, the level on which the country is interconnected to the rest of the world, and the variables of scale (ie.g.,  geographical and population sizes) into account\footnote{Data from the World Bank for the year 2011.}.
\newline
\newline
To use a diverse portfolio of countries in our analysis, we have obtained these variables for 188 countries and we have divided each of the five variables into quartiles of 47 countries each. Then, we  used an algorithm to create a choice of 70 countries so that, for each variable, we take approximately 70/4 (17 or 18) countries from each of the previously defined quartiles.
\newline
\newline
The algorithm that we use to create our database is rooted on a random search, and it works as follows. First, it considers the 188 countries and their position in each quartile out of the five variables that w selected. Then, it creates a random draw of 70 countries out of those 188. Afterwards, it computes how many times each quartile is represented in that random choice, and it then computes the standard deviation of that distribution. In a perfectly homogeneous distribution, each quartile should appear 87.5 times (70 countries times five variables divided by four quartiles) and the standard deviation should be 0. An example of a totally unequal choice is one in which the 70 countries chosen belong only to the first quartile of each variable.
\newline
\newline
Then, the algorithm draws another random choice and it again computes the standard deviation of the number of times each quartile is represented. If this is lower than the previous one, then it picks up the new choice. In contrast, if  it is higher, then it  keeps the former choice. This process is repeated 5000 times.
\newline
\newline
There is an additional constraint that affects the outcome of the final choice in this study: there are some countries that we believe should appear in our database regardless of the results of the algorithm---that is, the BRICS (Brazil, Russia, India, China and South Africa) and the United States---because of their increasing or consolidated geopolitical importance in the world.
\newline
\newline
We offer  a supplementary Excel file that contains a table which details the values of the five variables for each country chosen, it also gives a summary of the corresponding descriptive statistics and the source of each census used in our database.
\newline
\newline
Once a country is chosen, we define three criteria  to select the specific database that we will use in the empirical exercise: first, we take the latest available census; second, we consider (if possible) all of the population of the country, with no cut-off points regarding city size; and third, we take the smallest official administrative division as our urban unit, except when it is smaller than a city or village. We acknowledge that the equivalence of the urban units for such a large and heterogeneous number of countries is not attainable to the 100\%. Nevertheless, the application of the same three criteria for all of the considered nations allow us to think that the differences between the definitions of these units from country to country are minimized.
\newline
\newline
We have thoroughly analyzed each dataset, with reaggregation of certain spatial units when the division is more artificial than what we consider realistic in the official figures. For example, there are cases in which the dataset has the neighborhoods or subdistricts of a big city (usually the capital city) and small towns at the same spatial level. We aggregate the data of the neighborhoods of these big cities to create a unique urban unit. China and India were the countries that most needed these fine-tuning changes. For the case of the United States, we use the two most common datasets in the literature: the all places (unincorporated and incorporated) dataset for the year 2010 \citep{lude17,kwna19}, and the City Clustering Algorithm (CCA) Clusters for the year 2000 with a radius of 2 km \citep{RozRybGabMak11}.

\section{Results}
\label{Results}

\begin{figure}
\begin{tabular}{ccc}
\includegraphics[width=3.5cm]{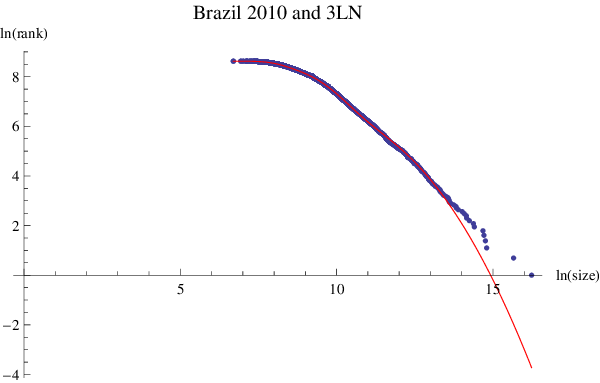}
&
\includegraphics[width=3.5cm]{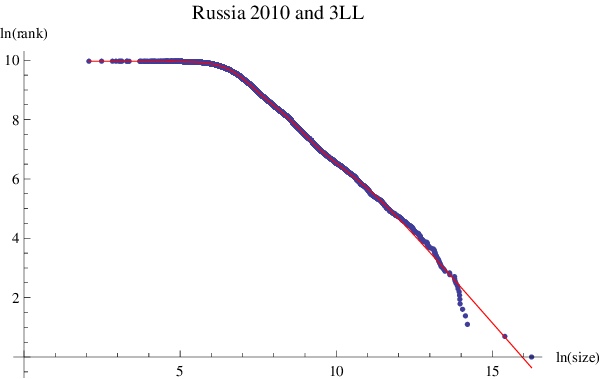}
&
\includegraphics[width=3.5cm]{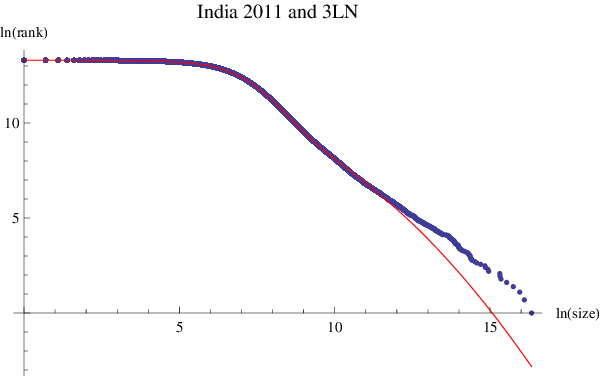}
\\
\includegraphics[width=3.5cm]{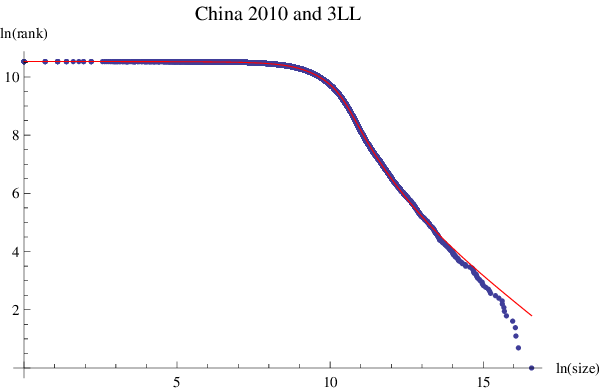}
&
\includegraphics[width=3.5cm]{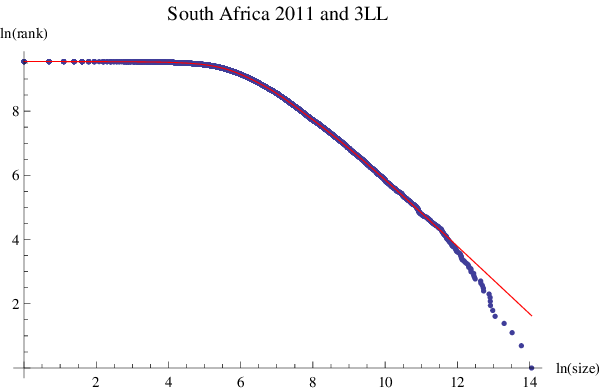}
&
\includegraphics[width=3.5cm]{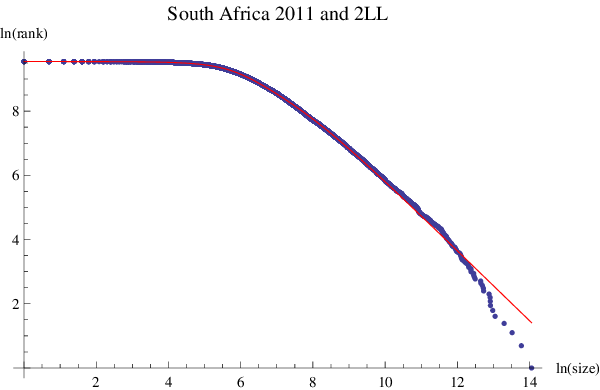}
\\
\includegraphics[width=3.5cm]{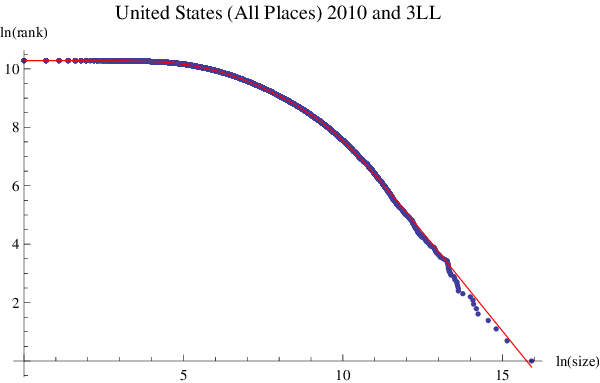}
&
\includegraphics[width=3.5cm]{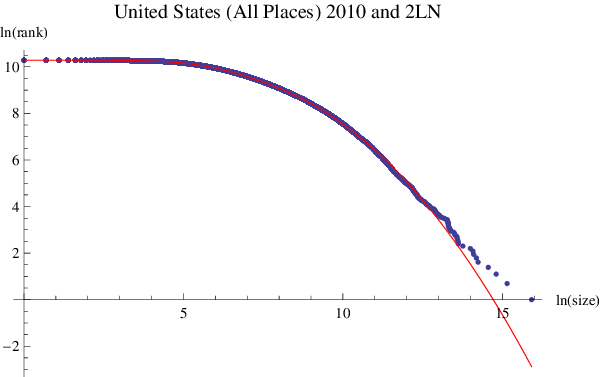}
&
\includegraphics[width=3.5cm]{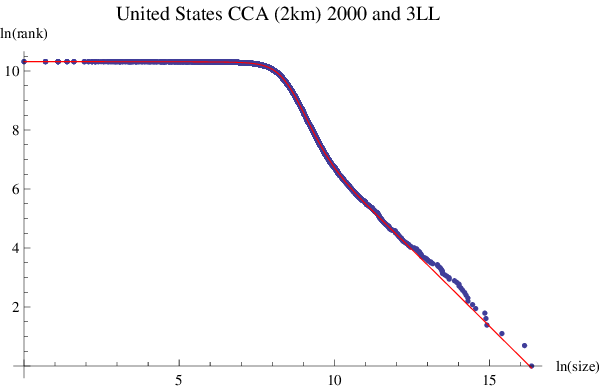}
\\
\end{tabular}
\caption{Rank-size log-log plots for several cases. In blue, the empirical data points; in red the predicted quantities according to the best model in each case.}
\label{figorda}
\end{figure}

In this paper we have estimated the parameters of the considered distributions by Maximum Likelihood Estimation (MLE). As an exception to the general rule, the log-likelihood function for the LN can be maximized in an exact way by the $\hat{\mu}$ equal to the sample mean of the logarithm of the data, and the $\hat{\sigma}$ the standard deviation of the logarithm of the data alike. In contrast, the log-likelihood functions for  LL, dPLN, 2LN, 2LL, 3LN and 3LL  cannot be maximized exactly\footnote{The log-likelihoods of the 2LN and 3LN can be maximized in a relatively simple way by an Expectation-Minimization (EM) algorithm (see, e.g, \cite{McLKri08}). However, the results are the same to the approach taken here up to, say, a precision of $10^{-4}$.} and one can resort to numerical methods.
We have used the commands included in the software {\sc MATLAB}$^{\circledR}$. Specifically,
we have used the commands {\tt mle} and {\tt mlecustom} that rely in turn on the command {\tt fminsearch}, that implements a multidimensional unconstrained nonlinear minimization (of minus the log-likelihood) according to the Nelder--Mead simplex method (direct search) \citep{LagReeWriWri98}.
We have treated in this way all of the distributions on an equal footing (for the LN the exact estimators and the numerical procedure give the same result up to an accuracy of, say, $10^{-4}$). This method requires the supply of initial values of the parameters to be estimated, which have been provided by choosing a suitable set of initial values and by trial-and-error. The iteration has two faces: each run of the algorithm repeats itself until the found values between two consecutive valuations are less than the prescribed tolerance (typically $10^{-4}$) or the maximum number of iterations are reached. Afterwards, in this last case, several runs of the algorithm may be necessary (supplying as initial values the outcome of the previous iteration). This is typically the case for mixture distributions because more iterations are required when there are more components.
\newline
\newline
Additionally, the Standard Errors (SE) of the ML estimators have been computed independently using {\sc Mathematica}$^{\circledR}$ software according to the indications of  \cite{McCVin03} and \cite{EfrHin78}.
This last reference advocates the use of observed Fisher information matrices for computing the standard errors of ML estimators, and this is the approach that we have followed.
We have also performed a simple $t$-ratio tests to assess the significance of the estimates so obtained. The results are presented, due to the length of the information provided, in three sheets of an Excel file that is offered as supplementary material. However, a brief summary follows. For the LN, the parameters can be estimated and the parameters are significative in 100\% of the possible instances. For the LL, a similar result holds. For the dPLN, the parameters cannot be estimated in 2.82\% of the instances and the parameters are not significative 1.41\% of the cases. For the 2LN, the parameters are not estimated in 1.41\% of the cases and the estimations are not significative in 2.82\% of the possibilities. The 2LL can be estimated in 100\% of the instances and the rate of not significant drops to 0.28\%. For the 3LN model, the parameters cannot be estimated in 1.41\% of the instances and the estimations become not significative 1.06\% of the time. For  3LL, it can be estimated that 100\% of the time but these estimations become non-significative in 2.11\% of the total possibilities. For the mixture models with three components, the non-significance of the parameters is often associated to a weight parameter of the mixture. This criterion may mean that a mixture model with two components might be more appropriate.
\newline
\newline
As an indication of the goodness-of-fit of the proposed distributions to the samples of our database, we have performed standard Kolmogorov-Smirnov (KS), Cram\'er-von Mises (CM) and Anderson-Darling (AD) tests using Monte Carlo resampling employing 350 synthetic datasets of the size of the original datasets (this ranges {}from 23,800 simulated values for the case of Malta to 210,626,850 ones for the case of India) and obtaining the approximate $p$-values out of the simulated distribution of the corresponding test's statistic.
We should mention that the values obtained by this procedure can vary a little from run to run but the outcome of the tests, namely rejection or non-rejection at the 5\% level of significance, seems to be robust to these small differences. We report our obtained values ($p$-values (statistics)) in a sheet of our Excel supplementary file, and the final outcome is given in Table~\ref{tab:1}. This table shows the results for the 70 countries, ordered by sample size from the lowest to the highest, and the seven distributions analyzed. ``Non reject'' means that none of the three tests rejects the corresponding distribution; ``reject'' (in red) is the opposite case: the three tests reject the distribution; finally ``mixed'' (in green) is associated with mixed evidence in favor of the distribution: one or two tests rejects or reject it while two or one, respectively, do/does not. There is a blank space for the cases of dPLN for Denmark and Israel, of 2LN for Nicaragua
and of 3LN for Timor-Leste, which happens because the densities cannot be estimated (as we have sketched earlier).
\newline
\newline
Out of the 71 cases (69 countries plus the two used samples of the United States) the percentage of rejections is as follows: LN (53.52\% of rejections), LL (39.44\%), dPLN (14.08\%), 2LN (11.27\%), 2LL (8.45\%), 3LN (0\%) and 3LL (0\%).  The LN, with more than 50\% of rejections, and the LL, with almost 40\%, are generally not good descriptions of the city size distributions of the considered countries. Both densities, the LN and the LL, are increasingly rejected as the sample size increases: from Romania, which has 3181 cities, onwards, they are always rejected. In some sense, the more observations you have, the more likely it is that a less flexible distribution would do worse\footnote{We have taken this idea from an anonymous referee. We thank them for such an appropriate observation.}. Also, the statistical tests that we employ have increasing power with sample size \citep{RazWah2011}. In contrast, the 3LN and the 3LL are excellent parametric descriptions of the city size distributions of our 70 countries according to this criterion because they are never rejected.
\newline
\newline
Now the important question is: which is the best distribution for each country? Table~\ref{tab:2} gives us an answer. The AIC and BIC information criteria show the preferred distribution for each country. The main result is that there is no single function that systematically dominates the others for all countries. The number of times that each density is chosen according to each criterion is presented in Table~\ref{tab:3}. The distribution chosen more often according to the AIC is, by far, the 3LN, followed by the dPLN and the 3LL. As is expected, the BIC introduces a higher penalization for models with more parameters, and this criterion ranks first the dPLN, followed by the LL.
Regarding the BRICS and the United States, the 3LN is the best distribution for Brazil and India, while the 3LL is the preferred one for Russia, China and United States (CCA).
For South Africa, the AIC selects 3LL. Meanwhile, the BIC selects 2LL. While  3LL and the 2LN are preferred for United States (all places data).
\newline
\newline
As a graphical complement to our analysis, we also offer
in Figure~\ref{figorda} some Zipf plots (rank-size plots in log-log scale) for the cases of the BRICS and the United States (all places and CCA). We have plotted the selected models according to both AIC and BIC. In case of different outcomes of these two criteria, we have plotted both instances for the corresponding sample (in the sense of the previous paragraph). The fit in all cases is remarkable, taking into account that the discrepancies at the upper tails of these plots are amplified by the fact of taking logarithms
(see, e.g., \cite{GonRamSan13}), and the fit is especially excellent for the United States samples.

\section{Conclusions}
\label{conclusions}

An update of the analysis of city size distribution for a representative number of countries (i.e., 70) has been carried out. Seven distributions are estimated: LN (lognormal), LL (loglogistic), dPLN (double Pareto lognormal), 2LN, 3LN (a mixture of two and three lognormals), 2LL and 3LL (a mixture of two and three loglogistics). Two main conclusions emerge.
\newline
\newline
Our first conclusion is as follows. The LL and, especially, the LN are rejected for a very significant number of countries while the 3LN and the 3LL are always non-rejected and, therefore, provide excellent descriptions of the city size distributions.  The fact that two very similar densities (3LN and 3LL) are never rejected for countries so heterogeneous raises the interesting hypothesis that the urban hierarchy of cities might share a basic common root with regard to the urbanization process, in spite of the evident historical divergences between nations.
\newline
\newline
Moreover, our heterogeneous sample contains two different types of country. First, developed countries, where urbanization and industrialization have gone hand-in-hand, giving rise to what is known as ``parallel urbanization'' \citep{henderson2003upa} and the urbanization process is over or stabilized. Second, developing countries, where the urbanization process is in its early stage and the phenomenon of “urbanization without growth” (\citealp{fay1999urbanization}; \citealp{jedwab2015urbanization}; \citealp{gollin2016urbanization}) might prevail. This first conclusion relates to the classical contribution of \cite{zelinsky1971hypothesis}, who describes the universal (applicable to all nations, with the only difference of the period in which it occurs) change from a rural, scattered and homogeneous landscape to another more concentrated, diversified, and hierarchical landscape; in other words, an urban landscape\footnote{In this line of reasoning, \cite{pumain2015multilevel} also emphasize that, apart from the time period and some non-essential peculiarities, the urbanization process of the BRICS resembles that of Europe and the United States.}.
\newline
\newline
Our second conclusion is that, according to the standard information criteria AIC and BIC, the dominance of the 3LN and the 3LL (which  is our first conclusion) is questioned or, at least, qualified. Indeed, AIC ranks first, by a great amount, then the 3LN, followed by the dPLN and the 3LL, while the BIC ranks first the dPLN followed by the LL.
This is related to the issue or danger of over-fitting by using theoretical parametric density functions with a high number of parameters. In our study, the considered mixture models with high number of components tend to fit the empirical datasets very well, leading to statistical tests of goodness-of-fit (almost) never rejected, but in terms of the valuable information that they contain, they might not be the best models, so we might fall in over-fitting (as mentioned earlier). In fact, in terms of standard information criteria, the single LN or LL is sometimes chosen by AIC (corresponding mainly to lower sample sizes) and the dPLN as well is sometimes chosen. Moreover, a non-negligible number of times the 2LN and 2LL are preferred to the 3LN and 3LL. In the case of the BIC, the contrast is more pronounced, ranking first the dPLN, still keeping valid to a great extent the main conclusions of \cite{giesen2010size} and \cite{val15}.

\linespread{1}

\begin{table}
\caption{Results of the statistical tests for each distribution and country}
\label{tab:1}       % Give a unique label
\begin{tiny}
\begin{tabular}{llllllllll}
\hline\noalign{\smallskip}
COUNTRY	& YEAR	& OBS	& LN	& LL	& dPLN & 2LN & 2LL & 3LN	& 3LL  \\
\noalign{\smallskip}\hline\noalign{\smallskip}
Malta & 2011 & 68 & non reject & non reject & non reject & non reject & non reject & non reject & non reject \\
New Zealand & 2006 & 74 & non reject & non reject & non reject & non reject & non reject & non reject & non reject \\
Iceland & 2011 & 76 & non reject & non reject & non reject & non reject & non reject & non reject & non reject \\
Timor-Leste & 2010 & 79 & non reject & non reject & non reject & non reject & non reject &  & non reject \\
Denmark & 2011 & 98 & \textcolor{green}{mixed} & non reject &  & non reject & non reject & non reject & non reject \\
Luxembourg & 2011 & 116 & non reject & non reject & non reject & non reject & non reject & non reject & non reject \\
Latvia & 2011 & 119 & non reject & non reject & non reject & non reject & non reject & non reject & non reject \\
Nicaragua & 2005 & 153 & non reject & non reject & non reject &  & non reject & non reject & non reject \\
South Korea & 2010 & 163 & \textcolor{green}{mixed} & non reject & non reject & non reject & non reject & non reject & non reject \\
Tonga & 2011 & 163 & \textcolor{green}{mixed} & non reject & non reject & non reject & non reject & non reject & non reject \\
Israel & 2008 & 170 & non reject & non reject &  & non reject & non reject & non reject & non reject \\
Kiribati & 2010 & 182 & non reject & non reject & non reject & non reject & non reject & non reject & non reject \\
Kazakhstan & 2009 & 200 & \textcolor{green}{mixed} & non reject & non reject & non reject & non reject & non reject & non reject \\
Slovenia & 2002 & 210 & non reject & non reject & non reject & non reject & non reject & non reject & non reject \\
Mauritania & 2013 & 210 & \textcolor{green}{mixed} & non reject & non reject & non reject & non reject & non reject & non reject \\
Estonia & 2011 & 226 & \textcolor{red}{reject} & non reject & non reject & non reject & non reject & non reject & non reject \\
Chad & 2009 & 252 & non reject & non reject & non reject & non reject & non reject & non reject & non reject \\
Sweden & 2011 & 290 & \textcolor{green}{mixed} & non reject & non reject & non reject & non reject & non reject & non reject \\
Myanmar & 2014 & 330 & \textcolor{red}{reject} & \textcolor{red}{reject} & non reject & non reject & non reject & non reject & non reject \\
Guatemala & 2002 & 331 & non reject & non reject & non reject & non reject & non reject & non reject & non reject \\
Mongolia & 2014 & 332 & \textcolor{red}{reject} & \textcolor{red}{reject} & non reject & non reject & non reject & non reject & non reject \\
Finland & 2011 & 336 & non reject & non reject & non reject & non reject & non reject & non reject & non reject \\
Chile & 2002 & 342 & non reject & non reject & non reject & non reject & non reject & non reject & non reject \\
Burkina Faso & 2006 & 351 & non reject & non reject & non reject & non reject & non reject & non reject & non reject \\
Sudan & 2008 & 362 & \textcolor{green}{mixed} & non reject & non reject & non reject & non reject & non reject & non reject \\
Venezuela & 2010 & 372 & non reject & non reject & non reject & non reject & non reject & non reject & non reject \\
Cyprus & 2011 & 388 & non reject & non reject & non reject & non reject & non reject & non reject & non reject \\
Iran & 2011 & 397 & \textcolor{red}{reject} & non reject & non reject & non reject & non reject & non reject & non reject \\
Rwanda & 2012 & 417 & non reject & non reject & non reject & non reject & non reject & non reject & non reject \\
Netherlands & 2011 & 418 & \textcolor{green}{mixed} & non reject & non reject & non reject & non reject & non reject & non reject \\
Norway & 2011 & 431 & non reject & non reject & non reject & non reject & non reject & non reject & non reject \\
Indonesia & 2010 & 492 & non reject & non reject & non reject & non reject & non reject & non reject & non reject \\
Lithuania & 2011 & 540 & \textcolor{red}{reject} & \textcolor{red}{reject} & non reject & non reject & non reject & non reject & non reject \\
Croacia & 2011 & 556 & \textcolor{red}{reject} & \textcolor{green}{mixed} & non reject & non reject & non reject & non reject & non reject \\
Belgium & 2011 & 571 & non reject & non reject & non reject & non reject & non reject & non reject & non reject \\
Ethiopia & 2007 & 731 & \textcolor{red}{reject} & \textcolor{green}{mixed} & non reject & non reject & non reject & non reject & non reject \\
Ireland & 2011 & 824 & \textcolor{red}{reject} & \textcolor{red}{reject} & non reject & non reject & non reject & non reject & non reject \\
Armenia & 2011 & 971 & \textcolor{green}{mixed} & non reject & non reject & non reject & non reject & non reject & non reject \\
Colombia & 2005 & 1113 & \textcolor{red}{reject} & non reject & non reject & non reject & non reject & non reject & non reject \\
Algeria & 2008 & 1484 & \textcolor{red}{reject} & non reject & non reject & non reject & non reject & non reject & non reject \\
Morocco & 2014 & 1499 & \textcolor{red}{reject} & \textcolor{green}{mixed} & non reject & non reject & non reject & non reject & non reject \\
Japan & 2010 & 1728 & non reject & non reject & non reject & non reject & non reject & non reject & non reject \\
Australia & 2011 & 1838 & \textcolor{red}{reject} & \textcolor{red}{reject} & \textcolor{green}{mixed} & \textcolor{red}{reject} & \textcolor{green}{mixed} & non reject & non reject \\
Austria & 2011 & 2357 & \textcolor{red}{reject} & non reject & non reject & non reject & non reject & non reject & non reject \\
Mexico & 2010 & 2456 & non reject & non reject & non reject & non reject & non reject & non reject & non reject \\
Poland & 2011 & 2479 & \textcolor{red}{reject} & \textcolor{red}{reject} & non reject & \textcolor{green}{mixed} & non reject & non reject & non reject \\
Switzerland & 2011 & 2515 & non reject & non reject & non reject & non reject & non reject & non reject & non reject \\
Slovakia & 2011 & 2889 & \textcolor{red}{reject} & non reject & non reject & non reject & non reject & non reject & non reject \\
Turkey & 2011 & 2934 & \textcolor{red}{reject} & \textcolor{red}{reject} & \textcolor{red}{reject} & non reject & non reject & non reject & non reject \\
Hungary & 2011 & 3154 & \textcolor{red}{reject} & non reject & non reject & non reject & non reject & non reject & non reject \\
Romania & 2011 & 3181 & \textcolor{red}{reject} & \textcolor{red}{reject} & non reject & non reject & non reject & non reject & non reject \\
Argentina & 2010 & 3430 & \textcolor{red}{reject} & \textcolor{red}{reject} & non reject & non reject & non reject & non reject & non reject \\
Nepal & 2011 & 4049 & \textcolor{red}{reject} & \textcolor{red}{reject} & non reject & non reject & non reject & non reject & non reject \\
Canada & 2011 & 4931 & \textcolor{red}{reject} & \textcolor{red}{reject} & non reject & non reject & non reject & non reject & non reject \\
Bulgaria & 2011 & 5121 & \textcolor{red}{reject} & \textcolor{red}{reject} & \textcolor{green}{mixed} & \textcolor{red}{reject} & non reject & non reject & non reject \\
Brazil & 2010 & 5565 & \textcolor{red}{reject} & \textcolor{red}{reject} & \textcolor{red}{reject} & \textcolor{red}{reject} & \textcolor{red}{reject} & \textcolor{green}{mixed} & non reject \\
Greece & 2011 & 6121 & \textcolor{red}{reject} & \textcolor{red}{reject} & \textcolor{red}{reject} & \textcolor{red}{reject} & \textcolor{green}{mixed} & non reject & non reject \\
Czech Republic & 2011 & 6251 & \textcolor{red}{reject} & \textcolor{red}{reject} & non reject & non reject & \textcolor{red}{reject} & non reject & non reject \\
Botswana & 2011 & 6610 & \textcolor{red}{reject} & \textcolor{red}{reject} & \textcolor{red}{reject} & \textcolor{green}{mixed} & \textcolor{red}{reject} & \textcolor{green}{mixed} & \textcolor{green}{mixed} \\
United Kingdom & 2011 & 7724 & \textcolor{red}{reject} & \textcolor{red}{reject} & \textcolor{red}{reject} & \textcolor{green}{mixed} & \textcolor{green}{mixed} & \textcolor{green}{mixed} & \textcolor{green}{mixed} \\
Italy & 2011 & 8094 & \textcolor{red}{reject} & \textcolor{red}{reject} & non reject & non reject & non reject & non reject & non reject \\
Spain & 2011 & 8115 & \textcolor{red}{reject} & \textcolor{red}{reject} & \textcolor{red}{reject} & non reject & non reject & non reject & non reject \\
Germany & 2011 & 11292 & \textcolor{red}{reject} & \textcolor{red}{reject} & \textcolor{red}{reject} & non reject & non reject & non reject & non reject \\
South Africa & 2011 & 13942 & \textcolor{red}{reject} & \textcolor{red}{reject} & \textcolor{green}{mixed} & non reject & non reject & non reject & non reject \\
Portugal & 2011 & 15151 & \textcolor{red}{reject} & \textcolor{red}{reject} & non reject & non reject & non reject & non reject & non reject \\
Russia & 2010 & 21354 & \textcolor{red}{reject} & \textcolor{red}{reject} & non reject & \textcolor{red}{reject} & \textcolor{green}{mixed} & non reject & non reject \\
United States (all places) & 2010 & 29239 & \textcolor{red}{reject} & \textcolor{red}{reject} & \textcolor{red}{reject} & non reject & non reject & non reject & non reject \\
United States (CCA) & 2000 & 30201 & \textcolor{red}{reject} & \textcolor{red}{reject} & \textcolor{red}{reject} & \textcolor{red}{reject} & non reject & non reject & non reject \\
France & 2012 & 36658 & \textcolor{red}{reject} & \textcolor{red}{reject} & \textcolor{green}{mixed} & non reject & \textcolor{red}{reject} & non reject & non reject \\
China & 2010 & 37358 & \textcolor{red}{reject} & \textcolor{red}{reject} & \textcolor{red}{reject} & \textcolor{red}{reject} & \textcolor{red}{reject} & non reject & non reject \\
India & 2011 & 601791 & \textcolor{red}{reject} & \textcolor{red}{reject} & \textcolor{green}{mixed} & \textcolor{red}{reject} & \textcolor{red}{reject} & non reject & non reject \\
\noalign{\smallskip}\hline
\end{tabular}
\end{tiny}
\end{table}

\begin{table}
\caption{Selected distributions for each country according to AIC and BIC}
\label{tab:2}       % Give a unique label
\begin{tiny}
\begin{tabular}{lll}
\hline\noalign{\smallskip}
COUNTRY	& AIC & BIC	 \\
\noalign{\smallskip}\hline\noalign{\smallskip}
Malta & 2LL & LN \\
New Zealand & 3LN & LL \\
Iceland & LN & LN \\
Timor-Leste & 2LN & 2LN \\
Denmark & 3LN & 2LN \\
Luxembourg & dPLN & LL \\
Latvia & 3LN & dPLN \\
Nicaragua & 2LL & LL \\
South Korea & 2LN & dPLN \\
Tonga & 2LL & 2LL \\
Israel & 3LN & 3LN \\
Kiribati & dPLN & LL \\
Kazakhstan & 2LN & dPLN \\
Slovenia & LL & LL \\
Mauritania & 3LN & LL \\
Estonia & 2LL & 2LL \\
Chad & 2LN & LL \\
Sweden & 2LL & dPLN \\
Myanmar & 3LN & dPLN \\
Guatemala & 2LN & LL \\
Mongolia & 3LL & dPLN \\
Finland & dPLN & LL \\
Chile & 3LL & 2LL \\
Burkina Faso & 2LN & LL \\
Sudan & dPLN & LL \\
Venezuela & LL & LL \\
Cyprus & LN & LN \\
Iran & dPLN & dPLN \\
Rwanda & LN & LN \\
Netherlands & 3LN & LL \\
Norway & dPLN & dPLN \\
Indonesia & 2LL & 2LL \\
Lithuania & 2LN & 2LL \\
Croacia & dPLN & dPLN \\
Belgium & 3LN & LL \\
Ethiopia & 3LN & dPLN \\
Ireland & dPLN & dPLN \\
Armenia & 2LL & 2LL \\
Colombia & 3LN & 2LL \\
Algeria & dPLN & dPLN \\
Morocco & 2LL & 2LL \\
Japan & 2LL & LN \\
Australia & 3LL & 3LL \\
Austria & 3LN & dPLN \\
Mexico & 3LN & 2LN \\
Poland & dPLN & dPLN \\
Switzerland & 2LN & LN \\
Slovakia & 3LN & 2LL \\
Turkey & 3LN & 3LN \\
Hungary & 3LL & dPLN \\
Romania & 3LN & 2LN \\
Argentina & 3LN & 2LN \\
Nepal & 3LL & dPLN \\
Canada & 3LN & 2LN \\
Bulgaria & 3LN & 3LN \\
Brazil & 3LN & 3LN \\
Greece & 3LL & 3LL \\
Czech Republic & 3LN & dPLN \\
Botswana & 3LN & 2LN \\
United Kingdom & 3LN & 3LN \\
Italy & dPLN & dPLN \\
Spain & 3LN & 2LN \\
Germany & 3LN & 3LN \\
South Africa & 3LL & 2LL \\
Portugal & 3LN & 2LN \\
Russia & 3LL & 3LL \\
United States (all places) & 3LL & 2LN \\
United States (CCA) & 3LL & 3LL \\
France & 3LN & 3LN \\
China & 3LL & 3LL \\
India & 3LN & 3LN \\
\noalign{\smallskip}\hline
\end{tabular}
\end{tiny}
\end{table}

\begin{table}
\caption{Summary of Table 2. Percentages of choice of the different density functions.}
\label{tab:3}       % Give a unique label
\begin{tabular}{lll}
\hline\noalign{\smallskip}
& AIC & BIC	 \\
\noalign{\smallskip}\hline\noalign{\smallskip}
LN & 3 (4.23\%) & 6 (8.45\%) \\
LL & 2 (2.82\%) & 14 (19.72\%) \\
dPLN & 11 (15.49\%) & 18 (25.35\%) \\
2LN & 8 (11.27\%) & 10 (14.08\%) \\
2LL & 9 (12.68\%) & 10 (14.08\%) \\
3LN & 27 (38.03\%) & 8 (11.27\%) \\
3LL & 11 (15.49\%) & 5 (7.04\%) \\
\noalign{\smallskip}\hline
\end{tabular}
\end{table}

\newpage
\clearpage

\newpage
\clearpage
%\begin{acknowledgements}
%\end{acknowledgements}

\bibliographystyle{spbasic}      % basic style, author-year citations
\bibliography{biblior2}   % name your BibTeX data base

% \begin{thebibliography}{}
% Author, Article title, Journal, Volume, page numbers (year)
% Author, Book title, page numbers. Publisher, place (year)
% etc
% \end{thebibliography}

\end{document}